\documentclass[12pt,a4paper]{article}

\usepackage{algorithm}
\usepackage{algpseudocode}
\usepackage{geometry}
\RequirePackage{amsmath,amssymb}

\RequirePackage{mathrsfs}
\RequirePackage{amsfonts}
\RequirePackage{dsfont}
\RequirePackage{braket}
\RequirePackage{tabularx}
\usepackage{multirow}
\RequirePackage{nicefrac}
\RequirePackage{graphicx}
\RequirePackage{booktabs}
\RequirePackage{colortbl}
\RequirePackage{xcolor}
\RequirePackage[symbol]{footmisc}
\usepackage{mathtools}
\usepackage{comment}
\usepackage{blkarray}
\usepackage{stackengine}
\usepackage{float}
\usepackage{rotating}
\usepackage{hhline}
\usepackage{tikz}
\usepackage{multirow}
\usepackage[section]{placeins}

\def\AEF{A.E. Faraggi}

\def\EJP#1#2#3{{\it Eur.\ Phys.\ Jour.}\/ {\bf C#1} (#2) #3}

\def\JHEP#1#2#3{{\it JHEP}\/ {\bf #1} (#2) #3}
\def\NPB#1#2#3{{\it Nucl.\ Phys.}\/ {\bf B#1} (#2) #3}
\def\PLB#1#2#3{{\it Phys.\ Lett.}\/ {\bf B#1} (#2) #3}
\def\PRD#1#2#3{{\it Phys.\ Rev.}\/ {\bf D#1} (#2) #3}
\def\PRL#1#2#3{{\it Phys.\ Rev.\ Lett.}\/ {\bf #1} (#2) #3}
\def\PRT#1#2#3{{\it Phys.\ Rep.}\/ {\bf#1} (#2) #3}
\def\etal{{\it et al\/}}

\def\beq{\begin{equation}}
\def\eeq{\end{equation}}
\def\beqn{\begin{eqnarray}}
\def\eeqn{\end{eqnarray}}

\usepackage{empheq}
\usepackage[most]{tcolorbox}
\usepackage{blkarray}
\newtcbox{\mymath}[1][]{%
    nobeforeafter, math upper, tcbox raise base,
    enhanced, colframe=blue!30!black,
    colback=blue!30, boxrule=1pt,
    #1}

\newcommand{\CC}[2]{C{#1\atopwithdelims[]#2}}

\newcommand{\ba}{\begin{eqnarray}}
\newcommand{\ea}{\end{eqnarray}}

\numberwithin{equation}{section}

\begin{document}
\begin{titlepage}
\samepage{
\setcounter{page}{1}
\rightline{}
\rightline{January 2021}

\vfill
\begin{center}
  {\Large \bf{
    Satisfiability Modulo Theories \\ \medskip and Chiral Heterotic String Vacua \\ \bigskip 
    with Positive Cosmological Constant}}

\vspace{1cm}
\vfill

{\large Alon E. Faraggi$^{1}$\footnote{E-mail address: alon.faraggi@liv.ac.uk}, 
 Benjamin Percival$^{1}$\footnote{E-mail address: benjamin.percival@liv.ac.uk}, \\ \medskip 
 Sven Schewe$^{2}$\footnote{E-mail address: svens@liv.ac.uk} and Dominik Wojtczak$^{2}$ \footnote{E-mail address: dkw@liv.ac.uk}
}
\\

\vspace{1cm}

{\it $^{1}$ Dept.\ of Mathematical Sciences, University of Liverpool, Liverpool
L69 7ZL, UK\\}
\vspace{.08in}
{\it $^{2}$ Dept.\ of Computer Sciences, University of Liverpool, Liverpool
L69 7ZL, UK\\}

\vspace{.025in}
\end{center}

\vfill
\begin{abstract}
\noindent

We apply Boolean Satisfiability (SAT) and Satisfiability Modulo Theories (SMT) solvers in the context of finding chiral heterotic string models with positive cosmological constant from $\mathbb{Z}_2\times \mathbb{Z}_2 $ orbifolds.
The power of using SAT/SMT solvers to sift large parameter spaces quickly to decide satisfiability, both to declare and prove unsatisfiability and to declare satisfiability, are demonstrated in this setting.
These models are partly chosen to be small enough to plot the performance against exhaustive search, which takes around 2 hours 20 minutes to comb through the parameter space. We show that making use of SMT based techniques with integer encoding is rather simple and effective, while a more careful Boolean SAT encoding provides a significant speed-up -- determining satisfiability or unsatisfiability has, in our experiments varied between 0.03 and 0.06 seconds, while determining all models (where models exist) took 19 seconds for a constraint system that allows for 2048 models and 8.4 seconds for a constraint system that admits 640 models.
We thus gain several orders of magnitude in speed, and this advantage is set to grow with a growing parameter space. This holds the promise that the method scales well beyond the initial problem we have used it for in this paper.

\end{abstract}

\smallskip}

\end{titlepage}

\section{Introduction}\label{intro}
Although the Standard Model (SM) 
of particle physics successfully predicts 
all observational data from particle collider experiments to date, 
several mysteries are left unanswered.
Among the most important of these mysteries are the origin of dark matter, how to stabilise the Higgs mass and how gravity may be incorporated into the framework.
To address these issues we can turn to String theory, which is the leading candidate for a theory beyond the Standard Model and naturally allows for the simultaneous analysis of gauge and gravitational interactions. 

The goal of identifying 4D string models that come as close as possible to the SM or 
Minimal Supersymmetric Standard Model (MSSM) has 
been pursued within a range of different approaches 
(see {\it e.g.} ref.~\cite{StringSM} and references therein). 
The main obstacle to these searches is the vast space of consistent string vacua in 4D, known as the string landscape. 
From a geometric perspective, this vast space of string vacua is generated by the enormous number of consistent compactifications of the 10D superstring down to 4D.
In this paper we work within the free fermionic worldsheet description~\cite{fff}
of the heterotic string, where the large space of vacua comes from the freedom in choosing the boundary conditions of worldsheet fermions.
This construction can be translated into the geometric context of $\mathbb{Z}_2\times \mathbb{Z}_2$ orbifolds as detailed in ref.~\cite{z2xz2}. 

The exploration of subspaces of the string landscape using Machine learning and Deep Learning has become a burgeoning area of research in string phenomenology, see, \textit{e.g.}, refs.~\cite{ML} and for review see, {\it e.g.}, ref.~\cite{MLReview}.
In this paper we make use of a different tool within this context:
Boolean Satisfiability Checking (SAT) and Satisfiability Modulo Theories (SMT). 
SAT and SMT solvers use powerful algorithms for determining the satisfiability of constraints as well as for finding solutions where the constraints are satisfiable.
We expect SAT and SMT solving to have a wide range of applications in string phenomenology as they have, indeed, proven to have in a range of other fields, such as computational biology~\cite{SMTbiology}, artificial intelligence \cite{SMTAI}, combinatorics \cite{SMTcombinatorics}, and
mathematical geometry \cite{SMTGeom}. 

Typically, the input variables needed to specify a string model are positive integers and \textit{a priori} there are approximately $100$ of them. 
They may, for example, specify the geometry of the compactified six-dimensional space or, in our case, generate boundary conditions for world-sheet free fermions. These inputs are required to meet certain consistency constraints, such as modular invariance, which reduces the number of independent input variables. In the case of free fermionic models, the boundary conditions are typically either Ramond or Neveu-Schwarz, and so the input variables can be represented as Boolean variables.
Classes of models can be analysed in terms of relations of these variables, and constraints related to desirable physical properties can be imposed.
Some examples of typical constraints are the presence of three particle generations, Higgs content and the absence of chiral exotic states. 
In recent works we have also been interested in unphysical, but interesting, string vacua, which are free of massless fermions or massless twisted bosons that are called Type 0 \cite{type0} and Type $\bar{0}$ \cite{type0bar} models, respectively.
These models have potential applications within cosmological scenarios of the early universe. 

As more and more phenomenological constraints are imposed, the frequency of viable models can become low enough to evade random searches that take many weeks.
This was found, for example, within the classification of Standard-Like Models and Left-Right Symmetric models \cite{slmclass,lrsclass} where, in the latter case, models would satisfy all imposed phenomenological criteria with a probability in the order of $10^{-11}$.
This scarcity of viable models has motivated the introduction of the fertility methodology within these classes of vacua \cite{slmclass,ferlrs}.
Furthermore, in some classes of vacua, important distinct characteristics of string models have been found to be incompatible.
For example, within the classification of $10^{12}$ Flipped $SU(5)$ vacua \cite{frs}, there were no exophobic models with three generations, whereas such vacua were found within the classification of Pati-Salam models \cite{acfkr}.
A similar result was found in recent work on Type 0 vacua~\cite{type0}, where it was shown analytically for a minimal class of models that the absence of massless fermions necessitates the presence of tachyons, and the result persisted within a generalised class of models.
As the construction of models becomes more complex and realistic, it is not straightforward to provide an analytic proof of apparent incompatibilities for phenomenological constraints that appear in classification statistics.
Scenarios of this kind are the strong suit of SAT and SMT, as they are powerful and proven tools in both finding rare solutions (and thus uncovering models with interesting properties) and proving unsatisfiability for constraints and isolating the origin of this unsatisfiability. 

In this paper we first of all test the efficiency of reducing the problem of finding tachyon-free Type $\bar{0}$ models of the sort found in \cite{type0bar} to SAT or SMT.
We first employ a standard SMT solver, Z3 \cite{Z3},
in an intuitive direct encoding of the input variables as integers, where all parameters are either 0 (corresponding to NS boundary conditions) or 1 (corresponding to R boundary conditions), which results in an efficiency improvement compared with a random classification: within the class of models we explore, the random classification approach finds approximately 500 tachyon-free Type $\bar{0}$ vacua in 40 minutes, which is the time it takes the SMT with integer encoding to find all 2048 solutions and to complete the search. 
We then go further to translate the constraints into a Boolean encoding such that the system is simply a SAT problem and the full power of the SAT solver, also part of Z3 \cite{Z3}, is demonstrated. In this case, establishing satisfiability takes 0.04 seconds while constructing all models takes 19 seconds, which is 126 times faster than with integer encoding, and around 450 times faster than enumerating and testing all candidate models.

Having established this, we then employ the SAT and SMT solvers to confirm a contradiction between the presence of spinorial $\mathbf{16}$'s for (tachyon-free) Type $\bar{0}$ models.
The SAT solver not only confirms the unsatisfiability of these constraints in 0.06 seconds in the Boolean encoding (163 seconds for the SMT solver using integer encoding), which is more than 100,000 times faster than an exhausting search. The SAT solver can also be used to identify a minimal `unsatisfiable core' \cite{UnsatCore}, which isolates where the contradiction arises by giving a (locally) minimal subset of constraints, where dropping either of them results in a satisfiable constraint system.
From this, it is straightforward to apply Optimisation Modulo Theory (OMT), offered by many SMT solvers, to find models with a minimal number of massless twisted bosons that also contain $\mathbf{16}/\overline{\mathbf{16}}$'s.
Where OMT is not offered, one can also add a minimisation/maximisation constraint to manually hone in on contradictions and optimal configurations.
Due to the abundance of massless fermions in their spectra, such models are expected to have a positive cosmological constant at the free fermionic point and could be candidates for having a necessarily positive one-loop potential once the string moduli are incorporated, as discussed in ref.\ \cite{FR}.  

The models we explore can be regarded as compactifications of the 
non-supersymmetric 10D tachyonic heterotic string
vacua \cite{spwsp,stable,so10tclass,PStclass}. By construction, the class of models we analyse here cannot give rise to (quasi-)realistic string vacua but rather is chosen as a simplified set-up, which is perfect for illustrating some key applications of SAT and SMT solvers.
We expect the advantages of the SAT/SMT approach to magnify for more realistic classes of models with larger input spaces.

The structure of the paper is as follows: in Section 2 we provide an introduction to SAT/SMT. In Section 3 we define the class of models we will explore and the constraints we evaluate. In Section 4 we detail how these can be evaluated using an SMT and SAT solvers and present the results. Section 5 concludes the paper.


\section{SMT and SAT}

Satisfiability Modulo Theories (SMTs) are powerful algorithms used for deciding whether a set of constraints describing a problem is satisfiable. In other words, SMTs determine whether there exists a `satisfying assignment' of a set of input variables to a system of constraints. These constraint formulae are constructed by defining operations over, what are referred to as, theory variables, and combining them with logical connectives. SMT problems are more expressive and powerful than the 
Boolean Satisfiability (SAT) problems that restrict all variables to be true or false and operators to be logical connectives. In particular, SMTs allow for operations over non-Boolean types such as integers, reals, bitvectors, and arrays. Both SMT and SAT are canonical NP-complete problems \cite{NP}, which is a class of computational problems for which checking whether a given variable assignment satisfies the constraints can be done in polynomial time, but finding such an assignment is believed to be hard. Despite a lot of effort, no polynomial time algorithm was shown for any NP-complete problem since the class was defined in 1971. More than that, a widely believed conjecture \cite{ETH} states that in the worst-case one cannot significantly improve over an exhaustive search of all possible assignments. Nevertheless, there has been a tremendous progress over the last decades in efficiency of algorithms solving these problems in practice, solving instances with hundreds and even thousands of variables, which shows that truly hard instances are few and far between.

A key aspect of how SMT-solvers work so effectively is through following the DPLL or 
conflict-driven clause learning (CDCL) class of algorithms. These algorithms implement a decision procedure for each theory by adding or subtracting constraints and querying for satisfiability as it goes. More detail on DPLL(T) and other decision procedures may be found in e.g. \cite{DPLL}.


One of the most efficient and easy to use SMT solvers is Z3, which can be found open source on Github at https://github.com/Z3Prover/Z3. Z3 was developed by Microsoft primarily for software verification purposes. It also implements an efficient SAT solver that we use for the Boolean encoding of our problem. It has bindings for most common programming languages and in our case we used the Python front-end as a means of interfacing with Z3. 


\section{Minimal Tachyon-free $SO(10)$ $\tilde{S}$-models 
}\label{Application}
We will utilise the free fermionic construction \cite{fff} to define a class of non-supersymmetric $\mathbb{Z}_2\times \mathbb{Z}_2$ orbifold models with an unbroken $SO(10)$ observable group. Only the key aspects of the free fermionic construction will be described here, as our main purpose is the application of the SMT solver within this setting. We will be adopting the conventional notation used in the free fermionic literature
\cite{fsu5, fny, alr, slm, lrs, acfkr, frs, FR, slmclass, lrsclass, ferlrs}.  
Over the past two decades
systematic methods to classify large numbers of free fermionic heterotic-string 
models were developed  \cite{fknr, fkr, acfkr, frs, slmclass, lrsclass, so10tclass, PStclass}. 
The initial method was developed for the classification of spinorial and anti-spinorial 
representations of an unbroken $SO(10)$ GUT group \cite{fknr}, and extended to include its
vectorial representations \cite{fkr}, which led to the discovery of spinor-vector duality 
over the space of vacua \cite{fkr,svd}. It was extended to include the entire 
massless twisted spectrum in models with, 
$SO(6)\times SO(4)$ \cite{acfkr},
$SU(5)\times U(1)$ \cite{frs}, 
$SU(3)\times SU(2)\times U(1)^2$ \cite{slmclass}, 
$SU(3)\times U(1)\times SU(2)^2$ \cite{lrsclass, ferlrs}, 
unbroken $SO(10)$ subgroups. Exophobic three generation models, in which exotic fractionally charged states only appear in the massive string spectrum were discovered in the case of
$SO(6)\times SO(4)$ models, whereas all other cases contained non-chiral massless exotic 
states in the spectra of three generation models. Over the past year
the classification methodology was extended to non-supersymmetric heterotic-string vacua
\cite{so10tclass, type0, PStclass, type0bar}, in which case the existence of 
physical tachyonic states in the physical spectrum is of particular interest. 
In the process of such classifications, we are particularly interested in the 
presence, or lack thereof, of some specific states in the spectrum of the models.
It is evident that SAT and SMT solving are particularly well suited to address such questions.

The construction of a free fermionic string model is defined at the free fermionic point in the moduli space and is generated by specifying two ingredients. The first is a set of $N$ boundary condition basis vectors, $v_i \in \mathcal{B}$, $i=1,...,N$, and the second is a set of one-loop Generalised
GSO (GGSO) phases, $\CC{v_i}{v_j}$. 
  
We will use the basis explored in \cite{type0bar}, where Type $\bar{0}$ string vacua, \textit{i.e.} models without any twisted massless bosons, were uncovered. This basis is written as
\begin{align}\label{basisStTi}
\mathds{1}&=\{\psi^\mu,\
\chi^{1,\dots,6},y^{1,\dots,6}, w^{1,\dots,6}\ | \ \overline{y}^{1,\dots,6},\overline{w}^{1,\dots,6},
\overline{\psi}^{1,\dots,5},\overline{\eta}^{1,2,3},\overline{\phi}^{1,\dots,8}\},\nonumber\\
\tilde{S}&=\{{\psi^\mu},\chi^{1,\dots,6} \ | \ \overline{\phi}^{3,4,5,6}\},\nonumber\\
{T_1}&=\{y^{1,2},w^{1,2}\; | \; \overline{y}^{1,2},\overline{w}^{1,2}\},\nonumber\\ 
{T_2}&=\{y^{3,4},w^{3,4}\; | \; \overline{y}^{3,4},\overline{w}^{3,4}\},\nonumber\\ 
{T_3}&=\{y^{5,6},w^{5,6}\; | \; \overline{y}^{5,6},\overline{w}^{5,6}\},\nonumber\\ 
{b_1}&=\{\psi^{\mu},\chi^{12},y^{34},y^{56}\; | \; \overline{y}^{34},
\overline{y}^{56},\overline{\eta}^1,\overline{\psi}^{1,\dots,5}\},\\
{b_2}&=\{\psi^{\mu},\chi^{34},y^{12},w^{56}\; | \; \overline{y}^{12},
\overline{w}^{56},\overline{\eta}^2,\overline{\psi}^{1,\dots,5}\},\nonumber\\
{b_3}&=\{\psi^{\mu},\chi^{56},w^{12},w^{34}\; | \; \overline{w}^{12},
\overline{w}^{34},\overline{\eta}^3,\overline{\psi}^{1,\dots,5}\},\nonumber\\
z_1&=\{\overline{\phi}^{1,\dots,4}\}.\nonumber
\nonumber
\end{align}
We note that the vectors $T_i$, $i=1,2,3$ allow for internal symmetric shifts around the 3 internal $T^2$ tori. The inclusion of $\tilde{S}$ in the basis means we have no supersymmetry and these vacua can be considered as compactifications of the tachyonic 10D heterotic string, as discussed in \cite{spwsp,stable,so10tclass, PStclass}. 

The NS sector vector gauge bosons give rise to a gauge group
\beq 
SO(10)\times U(1)^3\times SO(4)^3\times SU(2)^8
\eeq 
and may receive additional vector boson enhancements from the sectors
\beq \label{enhancements}
\begin{Bmatrix}
\psi^\mu\ket{z_1}_L \otimes \{\bar{\lambda^i}\}\ket{z_1}_R \\ 
\psi^\mu\ket{z_2}_L \otimes \{\bar{\lambda^i}\}\ket{z_2}_R \\
\psi^\mu \ket{z_1+z_2}_L \otimes \ket{z_1+z_2}_R
\end{Bmatrix}
\eeq  
where $\bar{\lambda}^i$ are all possible right moving Neveu-Schwarz oscillators and $z_2$ is the important linear combination
\beq 
z_2=1+b_1+b_2+b_3+z_1=\{\bar{\phi}^{5,6,7,8}\}.
\eeq
Having defined these 9 basis vectors, the other ingredient for defining a model is the GGSO coefficients, $\CC{v_i}{v_j}=\pm 1$, which will generate a $9\times 9$ matrix for this basis but, due to modular invariance constraints, only the upper triangle of 36 GGSO phases are independent. This leaves a space of $2^{36}\sim 6.87 \times 10^{10}$ independent GGSO phase configurations.

With an assignment of GGSO phases the Hilbert space of states $\ket{S_\xi}$ can be specified through 
\begin{equation}\label{hilbert}
    \mathcal{H}=\bigoplus_{\xi\in\Xi}\prod^{k}_{i=1}
\left\{ e^{i\pi v_i\cdot F_{\xi}}\ket{S_\xi}=\delta_{\xi}
\CC{\xi}{v_i}^*\ket{S_\xi}\right\}. 
\end{equation}
where $\xi$ are sectors (linear combinations of basis vectors) in the additive space $\Xi$ and $\delta_{\xi}$ is the spin statistic index. The GGSO projection equation inside the curly brackets of (\ref{hilbert}) will give us the constraints implemented within the SMT solver described in the next section.

The sectors in the model can be characterised according to their left and
right moving vacuum separately
\begin{align}
M_L^2&=-\frac{1}{2}+\frac{\xi_L \cdot\xi_L}{8}+N_L\\
M_R^2 &=-1+\frac{\xi_R \cdot\xi_R}{8}+N_R \nonumber
\end{align}
where $N_L$ and $N_R$ are sums over left and right moving oscillators, 
respectively. Physical states must additionally satisfy the
Virasoro matching condition, $M_L^2=M_R^2$ such that massless states are those with $M_L^2=M_R^2=0$ and physical tachyons arise for sectors with $M_L^2=M_R^2<0$. 

\subsection{Massless Bosonic Sector Analysis}
The first aspect of these models we wish to explore is the massless twisted bosons. In \cite{type0bar}, we have described the conditions on the absence of twisted massless bosons within this class of models and found tachyon-free GGSO configurations that corresponded to two distinct partition functions. The details are repeated here as we will apply the SMT solver to these conditions on the absence of massless twisted bosons.  

For this class of models, there are 15 vectorial bosonic sectors of the form
\begin{align}
    \begin{split}\label{VectBosonsTi}
       V^1_{pq}&=b_2+b_3+T_1+pT_2+qT_3\\ 
       V^2_{pq}&=b_1+b_3+T_2+pT_1+qT_3\\ 
       V^3_{pq}&=b_1+b_2+T_3+pT_1+qT_2\\
       V^4&=T_1+T_2 \\
       V^5&=T_1+T_3 \\
       V^6&=T_2+T_3 
    \end{split}
\end{align}
where $p,q=0,1$ and these sectors all come with a right-moving NS oscillator. Additionally, there are 30 fermionic spinorial sectors
\begin{align}
    \begin{split}\label{FermBosonsTi}
       B^1_{pq}&=b_2+b_3+z_1+T_1+pT_2+qT_3\\ 
       B^2_{pq}&=b_1+b_3+z_1+T_2+pT_1+qT_3\\ 
       B^3_{pq}&=b_1+b_2+z_1+T_3+pT_1+qT_2 \\
       B^4_{pq}&=\mathds{1}+b_1+z_1+T_1+pT_2+qT_3\\ 
       B^5_{pq}&=\mathds{1}+b_2+z_1+T_2+pT_1+qT_3\\ 
       B^6_{pq}&=\mathds{1}+b_3+z_1+T_3+pT_1+qT_2.\\ 
       B^7&=T_1+T_2+z_1\\
       B^8&=T_1+T_3+z_1 \\
       B^9&=T_2+T_3+z_1\\ 
       B^{10}&=T_1+T_2+z_2\\
       B^{11}&=T_1+T_3+z_2 \\
       B^{12}&=T_2+T_3+z_2\\ 
    \end{split}
\end{align}
A Type $\bar{0}$ model arises when all of these sectors are projected for an assignment of GGSO phases. 
For example, taking sector $B^1_{pq}$, a projector is constructed of the form
\beq \label{B1proj}
P_{pq}^1=\frac{1}{2^3}\left(1+\CC{B^1_{pq}}{T_1}\right)\left(1+\CC{B^1_{pq}}{z_2}\right)\left(1+\CC{B^1_{pq}}{b_1+pT_2+qT_3}\right)
\eeq
and requiring that this is zero ensures its absence from the massless spectrum. Repeating this for all massless twisted bosonic sectors allows for the identification of Type $\bar{0}$ configurations.
\subsection{Tachyon Sector Analysis}
The next constraint to impose after the projection of massless bosonic sectors, is the absence of tachyonic sectors. Since our models are non-supersymmetric this is an important check for determining the stability of our models for a 4D Minkowski background. The same procedure of encoding the GGSO projections applies to the tachyonic sectors. The tachyonic sectors in this construction are: $\{\bar{\lambda}\}T_i, z_1, z_2, z_1+T_i$ and $z_2+T_i$, $i=1,2,3$, where $i=1,2,3$ and $\bar{\lambda}$ is some right-moving NS oscillator and we note that the untwisted tachyon from $\ket{0}_L\otimes \{\bar{\lambda}\}\ket{0}_R$ is projected regardless of the GGSO phase choices in this basis. 

As an example, we can delineate the condition for the projection of the $\{\bar{\lambda}\}T_1$ tachyonic sector 
\beq 
\#(x\in S \ | \ x=-1)>1 \ \ \text{ where  } \ S=\left\{\CC{T_1}{T_2},\CC{T_1}{T_3},\CC{T_1}{z_1},\CC{T_1}{z_2}\right\}.
\eeq
which ensures all oscillator cases are projected.
\subsection{Spinorial $\mathbf{16}/\overline{\mathbf{16}}$ Sectors}
The fermion generations transforming in the spinorial $\mathbf{16}/\overline{\mathbf{16}}$ of $SO(10)$ arise from the twisted sectors
\begin{align}\label{spin16s}
F^1_{pq}&=b_1+pT_2+qT_3\nonumber\\ 
F^2_{pq}&=b_2+pT_1+qT_3\\
F^3_{pq}&=b_3+pT_1+qT_2\nonumber.
\end{align} 
which have projectors
\begin{align}\label{RProjs}
   R_{pq}^1=\frac{1}{2^3}\left(1-\CC{F^1_{pq}}{T_1}\right)\left(1-\CC{F^1_{pq}}{z_1}\right)\left(1-\CC{F^1_{pq}}{z_2}\right) \nonumber \\
   R_{pq}^2=\frac{1}{2^3}\left(1-\CC{F^2_{pq}}{T_2}\right)\left(1-\CC{F^2_{pq}}{z_1}\right)\left(1-\CC{F^2_{pq}}{z_2}\right) \\
   R_{pq}^3=\frac{1}{2^3}\left(1-\CC{F^3_{pq}}{T_3}\right)\left(1-\CC{F^3_{pq}}{z_1}\right)\left(1-\CC{F^3_{pq}}{z_2}\right) \nonumber 
\end{align}
which can be used to tell us $\#(\mathbf{16}+\overline{\mathbf{16}})$ for any model. Knowing this is sufficient for our purposes here with the SMT solver analysis and we will not implement the chirality projection distinguishing the $\mathbf{16}$ and $\overline{\mathbf{16}}$. 

\section{Application of SMT}

Now we turn our attention to the analysis of these model characteristics with an SMT written using Z3 in Python. It is convenient to introduce the notation 
\beq 
\CC{v_i}{v_j}=\exp{[i\pi( v_i | v_j )]}
\eeq 
and use the 36 independent phases $( v_i | v_j )\in \{0,1\}$ as the input variables for our SMT solver.
In terms of these variables, the GGSO projection equation for each sector can be written in terms of sums of $( v_i | v_j )$ modulo 2.
We call this representation the integer encoding for our system of constraints. 

As mentioned in Section 2, we use the general purpose SMT solver Z3, which allows us to use mathematical expressions so we do not need to reduce the constraints to purely propositional logic as in this integer encoding of the GGSO equations.
This comes at the expense of performance because the SMT solver needs to include reasoning for mathematical theories (such as integer arithmetic).
Therefore we will also detail a Boolean encoding of our constraint system, where we rewrite the GGSO projections purely as Boolean propositions. 

It turns out that the conditions for the projection of massless twisted bosons and tachyonic sectors do not involve the following 9 of the 36 $( v_i | v_j )$ phases:
\beq
(\mathds{1}|\tilde{S}), (\mathds{1}|b_1), (\mathds{1}|b_2), (\mathds{1}|b_3), (\mathds{1}|z_1), (\tilde{S}|b_1),(\tilde{S}|b_2),(\tilde{S}|b_3),(\tilde{S}|z_1)
\eeq 
and so our space of models to explore is reduced to $2^{27}$ (ca.\ $1.34 \times 10^{8}$). This is well within the reach of a complete enumeration of possible GGSO configurations but our purpose here is testing the features and efficiency of the SMT solver within this simple class of models. 
Moreover, as it does not adversely effect performance, we have re-introduced these 9 variables as input to our Boolean encoding with the expectation that it will not significantly impact the running time (in face, it has proven to actually cut the running time by 4\%.) 

As an example of how we can write the GGSO projections in both the integer representation and the Boolean representation we will take the massless bosonic sector $B^1_{00}$. The GGSO projection equation is written for generic $p,q=0,1$ in terms of $\CC{\cdot}{\cdot}$'s in eq. \ref{B1proj}. In the integer encoding we can write the $B^1_{00}$ projection condition by first defining
\begin{align}
P&=\left[(T_1|b_2)+(T_1|b_3)+(T_1|z_1)+1+(\mathds{1}|T_1)\right] \text{ mod } 2 \nonumber \\
Q&=\left[(\mathds{1}|T_1)+(T_1|b_1)+(T_1|b_2)+(T_1|b_3)+(T_1|z_1)+(b_1|b_2)+(b_1|b_3)+(b_1|z_1)\right]\text{ mod } 2 \nonumber \\
R&=\left[1+(b_1|b_2)+(b_1|b_3)+(b_1|z_1)+(T_1|b_1)\right] \text{ mod } 2 
\end{align}
and then constructing the conjunction
\begin{align}
     ( P= 1 \lor Q= 1  \lor R=1).
\end{align}
to impose the projection constraint.
This can then be translated into the Boolean representation using the \texttt{Xor()} operator, $\veebar$, as follows:
\begin{align}\label{BoolProj1}
\tilde{P}&=(T_1|b_2)\veebar(T_1|b_3)\veebar(T_1|z_1)\veebar \text{True}\veebar \mathds{1}|T_1) \nonumber \\
\tilde{Q}&=(\mathds{1}|T_1)\veebar(T_1|b_1)\veebar(T_1|b_2)\veebar(T_1|b_3)\veebar(T_1|z_1)\veebar(b_1|b_2)\veebar(b_1|b_3)\veebar(b_1|z_1) \nonumber \\
\tilde{R}&=\text{True}\veebar(b_1|b_2)\veebar (b_1|b_3) \veebar(b_1|z_1)\veebar(T_1|b_1) 
\end{align}
where $(v_i|v_j)$ are now Booleans rather than 0 or 1. The constraint for ensuring the projection of this sector is then
\begin{align}
     ( \tilde{P} \lor \tilde{Q}  \lor \tilde{R}).
\end{align}

For both representations we can schematically write the steps for the construction of the Z3 solver in the case of finding tachyon-free Type $\bar{0}$ string models\footnote{Both the integer and Boolean Z3 codes are available at https://github.com/thePlumbaked/SMTsType0bar.}
\begin{algorithm}
  \begin{algorithmic}[1]
     \State Define the 27 input variables $c_0,...,c_{26}$
      \State Add constraints on input variable domain $(c_i=0 \lor c_i= 1)$ $\forall i=0,...,26$.  
      \State Add constraints for GGSO projection of all twisted massless bosons
      \State Add constraints for GGSO projection of all tachyonic sectors.
      \State Check satisfiability OR
      \State Find satisfying assignments (print all solutions)
  \end{algorithmic}
\end{algorithm}
\FloatBarrier
Step 5 is perhaps the most fundamental application of SAT/SMT solvers, which can be used to quickly identify whether the set of constraints permits a solution or not. In this case, we can utilise the prior analysis in ref.~\cite{type0bar}, which found tachyon-free Type $\bar{0}$ models, to realise that the SMT solver should return \texttt{sat} in this case.

\subsection{Results of SMT search for tachyon-free Type $\bar{0}$ models}
Enumerating all tachyon-free Type $\bar{0}$ models within the class of models under consideration is a good testing ground for the efficiency of the SMT/SAT solver. As mentioned earlier, the space of models is $2^{27}$ (ca.\ $1.34 \times 10^8$), which is within the grasp of a complete enumeration approach.
We have run a random search in comparison, which is able to analyse ca.\ 16,000 sample points per second. An exhaustive enumeration of the model thus takes around 2 hours 20 minutes, and a random search (with repetitions) needs on average around 19 hours to find all solutions and 4 seconds to find the first model.

\begin{figure}[!htb]
\centering
\includegraphics[width=0.8\linewidth]{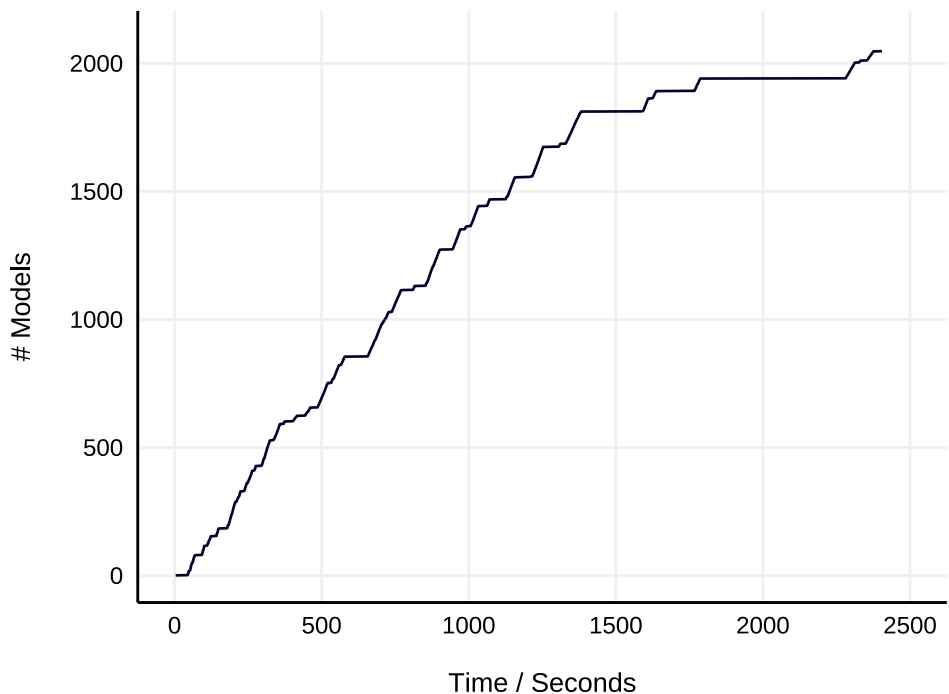}
\caption{\label{IntTimings}\emph{Rate at which the Integer representation finds all tachyon-free Type $\bar{0}$ models.}}
\end{figure}

Within the integer representation, Z3's SMT solver determines satisfiability in 12 seconds and finds all 2048 solutions in the full $2^{27}$ space in $2405$ seconds (ca.\ $40$ minutes). A random search found just under a quarter of these models in the same amount of time (500 models).
While this is a useful speed up, it is not that impressive. This is expected since, in this representation, the SMT solver deals with mathematical theories that create a significant overhead.
Figure \ref{IntTimings} depicts the accumulation of solutions over time for the SMT solver within the integer representation.

The performance is significantly improved when implementing a Boolean encoding. In this case, Z3's SAT solver determines satisfiablitiy in 0.04 seconds. It can find and print all 2048 solutions in 19 seconds.
This is 126 times faster than for the integer representation, and 450 times faster than exhaustively enumerating and checking the statespace.
\begin{figure}[!htb]
\centering
\includegraphics[width=0.8\linewidth]{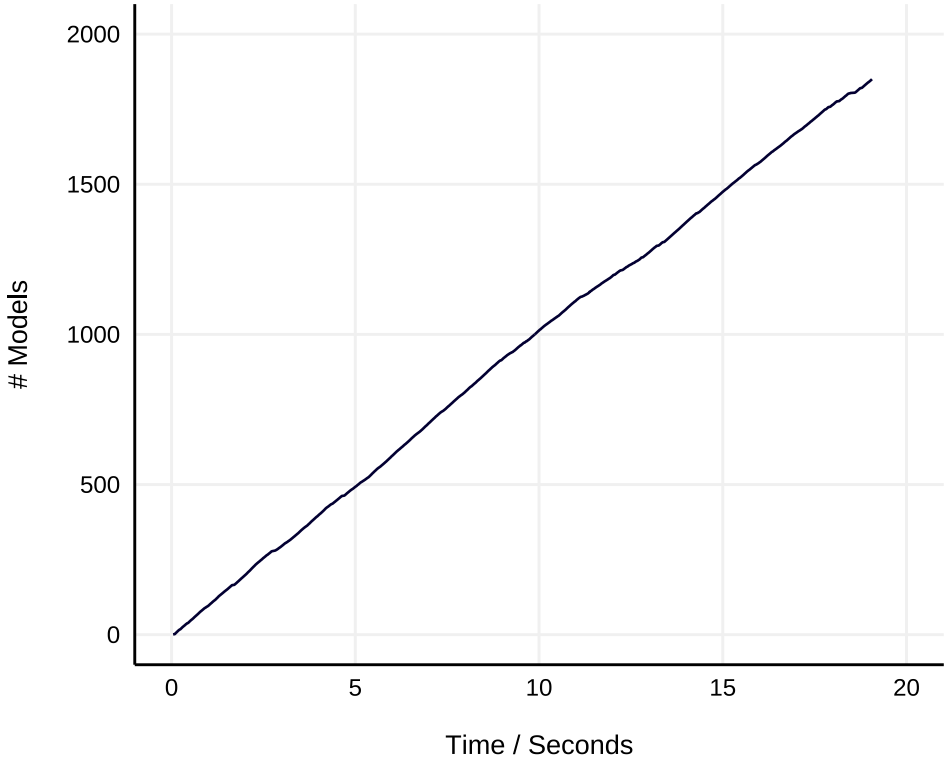}
\caption{\label{BoolTimings}\emph{Rate at which the Boolean representation finds all tachyon-free Type $\bar{0}$ models, depicting the number of compact solutions found; the 1850 compact solutions contain all 2048 explicit solutions (exactly once).}}
\end{figure}
Figure \ref{BoolTimings} depicts the accumulation of solutions over time for the SAT solver,
where we see that the solver is not slowed down by the creation of new lemmas as solutions are enumerated.
It can be seen that only 1850 compact solutions are recorded on the graph in Figure \ref{BoolTimings}. This is because some of the compact solutions include variables labelled with \texttt{None}, meaning that this variable/s can be set to true or false; they can be trivially expanded to the 2048 solutions (without omissions and without multiple occurrences of individual solutions).

We have repeated the experiment using all 36 original input variables.
Our expectation was that the SAT solver would essentially ignore them, as they do not enter into the reasoning at any point, though it will need to output a few additional \texttt{None} in the compact representation. We expected an insignificant increase in the overall running time.
We found that the solver worked 4\% faster, which is likely due to different decisions made by the heuristics.
But it shows an interesting effect:
overlooking the variables that do not matter did not slow the SAT solver down, whereas an exhaustive search would have taken 6 orders of magnitude longer.
\subsection{Identifying Chiral, Tachyon-free Type $\bar{0}$ Vacua}
In the analysis from ref.~\cite{type0bar} it was found that no Type $\bar{0}$ vacua include the fermion generations from the spinorial $\mathbf{16}/\overline{\mathbf{16}}$ sectors (\ref{spin16s}). Since these sectors are phenomenologically desirable, we aim to ensure at least one remains in the Hilbert space after GGSO projections. 

The sectors giving rise to the $\mathbf{16}/\overline{\mathbf{16}}$ were given in eq.\ (\ref{spin16s}) and the projectors of eq.~(\ref{RProjs}) can be rewritten in the integer and Boolean representations in a similar way as delineated above for $B^1_{00}$.

With the addition of this condition the SMT structure summary can be updated to
\begin{algorithm}
  \begin{algorithmic}[1]
     \State Define the 27 input variables $c_0,...,c_{26}$
      \State Add constraint on input variable domain $(c_i=0 \lor c_i= 1)$ $\forall i=0,...,26$.  
      \State Add constraints for GGSO projection of all twisted massless bosons
      \State Add constraints for GGSO projection of all tachyonic sectors.
      \State Add constraint on presence of at least 1 $\mathbf{16}/\overline{\mathbf{16}}$ sector
      \State Check satisfiability OR
      \State Find satisfying assignments (solutions)
  \end{algorithmic}
\end{algorithm}
\FloatBarrier
As expected from the findings of ref.\ \cite{type0bar}, with this added constraint on the spinorial $\mathbf{16}/\overline{\mathbf{16}}$ the Z3 solver returns \texttt{unsat}. This takes only 0.06 seconds in the Boolean encoding (163 seconds in the integer encoding), underlining the feasibility of combing large parameter spaces.

In such cases where constraint systems are unsatisfiable there are several tools within Z3 that are helpful to understand and isolate where the inconsistency arises from. Using its \texttt{proof()} method, Z3 will output a proof of inconsistency.
Unsatisfiability proofs can be long and tedious because, while satisfiability can be shown by providing a model, unsatisfiability needs to make a mathematical argument that establishes the contradiction.
Such proofs are often long and tedious---though this is no comparison to the tedious work of manually distilling the contradiction, but it is fair to say that they often do not provide much accessible further insight. 

There are, however, several other helpful tools that can be used to isolate inconsistencies. In particular, a minimal `unsatisfiable core' of constraints can be returned by most SMT solvers.
Additionally, a more manual approach of using push and pop methods on constraints allows for pinning down the source of an inconsistency.
Using this approach, the presence of the spinorial $\mathbf{16}/\overline{\mathbf{16}}$'s from $F^i_{pq}$ is found to contradict the projection of the vectorial $V^i_{pq}$ (under the presence of the remaining constraints). Since the Higgs bidoublet representation would reside within the vectorial $\mathbf{10}$ of $SO(10)$ coming from the $V^i_{pq}$, $i=1,2,3$, there is physical motivation to keep at least one of these twisted bosonic sectors in the Hilbert space.
Demanding that at least one sector from $F^1_{pq}$ remains, and at least one of $V^1_{pq}$ makes the SMT return \texttt{sat}, which takes 0.03 seconds for the Boolean constraint system and 8.4 seconds for the integer encoding.
Creating all 640 satisfying assignments in 7.5 seconds for the Boolean constraint system (1661 seconds for the integer encoding) of the 27 input variables with these conditions. One such model is defined by
{\begin{equation}
\small
\CC{v_i}{v_j}= 
\begin{blockarray}{ccccccccccc}
&\mathbf{1}& \tilde{S} &T_1&T_2&T_3& b_1 & b_2&b_3&z_1 \\
\begin{block}{c(rrrrrrrrrr)}
\mathbf{1}& 1& 1& 1& -1& 1& 1& -1& -1& 1 \\
\tilde{S} & 1& -1& -1& -1& -1& 1& 1& 1& 1 \\
T_1&  1& -1& -1& 1& 1& -1& 1& -1& 1\\
T_2& -1& -1& 1& 1& 1& -1& 1& 1& 1\\
T_3&  1& -1& 1& 1& -1& 1& 1& 1& 1\\
b_1&  1& -1& -1& -1& 1& 1& -1& -1& -1 \\
b_2& -1& -1& 1& 1& 1& -1& -1& -1& 1 \\
b_3& -1& -1& -1& 1& 1& -1& -1& -1& 1&\ \\
z_1&   1& -1& 1& 1& 1& -1& 1& 1& 1 \\ 
\end{block}
\end{blockarray}
\label{ggsophases}
\end{equation}}
which has partition function
\begin{equation}\label{PF}
     Z = 2\bar{q}^{-1} +1280q^{-1/2}\bar{q}^{1/4}+48q^{1/4}\bar{q}^{-3/4} -1016 - 16288 q^{1/4}\bar{q}^{1/4}  + \cdots
\end{equation}
where we can see that $N_b^0-N_f^0=-1016$, which is a large abundance of massless fermions as expected. This model generates a worldsheet cosmological constant of $\Lambda=-886.43$ which corresponds to a large positive cosmological constant, $\lambda$, via the rescaling $\lambda =  -\frac{1}{2}\mathcal{M}^4\Lambda$, where $\mathcal{M}=M_{String}/2\pi$.

\section{Conclusion}
The application of Machine Learning techniques within the string landscape is already a burgeoning field. In this work, we open the door to the application of SAT and SMT solvers within this context and have demonstrated their power and efficiency within a simplified class of string models and expect their benefits to only increase as more generalised and (quasi-)realistic classes of vacua are studied. 

We have demonstrated how SAT and SMT solvers can be used to help isolate inconsistent constraints and be used to optimise desired characteristics. This method was then employed to find string vacua with positive cosmological constant and desirable $SO(10)$ representations. Furthermore, this approach essentially meant minimising the number of massless twisted bosons, which is generally enough to ensure a large abundance of massless fermions resulting in the largest contribution to cosmological constant (the massless level) being positive. Although the massive and off-shell contributions also need accounting for, a large abundance of massless fermions can effectively guarantee a positive cosmological constant. Such vacua are then ripe for analysis away from the free fermionic point in the moduli space by exploring the one-loop potential in terms of the moduli fields as was pursued in \cite{FR}, where a model with $N_f>N_b$ at the free fermionic point and $N_b=N_f$ at a generic point in the moduli space resulted in guaranteeing a positivity of the one-loop potential. Using the SAT and SMT solvers described in the current work makes it very quick and easy to find string models at the free fermionic point with desired properties at the massless level, even when the models are rare in the full space.

There are a couple of obvious limitations, from a phenomenological perspective, in the class of models that we examined in this paper.
First of all, a $F^1_{pq}$ and $V^1_{pq}$ could not give rise to desired coupling as they reside on the same orbifold plane.
Secondly, due to only employing the $T_i$, $i=1,2,3$, in the basis we do not allow for shifts around the 6 circles of the internal $T^6$ and
there is a multiplicity factor of 4 attached to each sector, making 3 generation models impossible.
However, our focus here is on the introduction of SAT and SMT solving and the illustration of its application, while the application of SAT and SMT solving to more realistic constructions is left for future work.
It is evident, however, that as we seek to construct satisfiability criteria for non-supersymmetric \cite{nonsusy} and more detailed
phenomenological string models, for which the SAT and SMT algorithms are particularly well suited, these algorithmic approaches are of immense interest and utility.

\section*{Acknowledgments}

We would like to thank Viktor Matyas for helpful discussions. 
The work of BP is supported in part by STFC grant ST/N504130/1. The work of SS and DW are supported in part by EPSRC grant EP/P020909/1.

\bigskip

\bibliographystyle{unsrt}

\end{document}